\documentclass[mathleft]{an}
\usepackage{graphicx}
\usepackage{amsmath}
\usepackage{times}
\usepackage{amssymb}
\usepackage{hyperref}
\usepackage{breakurl}
\usepackage[T1]{fontenc}	
\makeatletter
\let\NAT@parse\undefined
\makeatother

\usepackage[numbers]{natbib}
\bibpunct{(}{)}{;}{n}{}{,}

\begin{document}

\Pagespan{1}{}
\Yearpublication{00}%
\Yearsubmission{00}%
\Month{0}%
\Volume{0}%
\Issue{0}%

\title{A fully automated data reduction pipeline for the FRODOSpec integral field spectrograph}

\author{
	R.M.Barnsley\inst{1}\thanks{rmb@astro.livjm.ac.uk} 
	, R.J.Smith\inst{1}
	\and I.A.Steele\inst{1}
}
\titlerunning{An automated pipeline for FRODOSpec}
\authorrunning{Barnsley, Smith \& Steele}
\institute{Astrophysics Research Institute, Liverpool John Moores University, Twelve Quays House, Birkenhead, CH41 1LD, U.K.
}



\abstract{A fully autonomous data reduction pipeline has been developed for FRODOSpec, an optical fibre-fed integral field spectrograph 
  currently in use at the Liverpool Telescope. This paper details the process required for the reduction of data taken using an integral field spectrograph
  and presents an overview of the computational methods implemented to create the pipeline. Analysis of errors and possible future enhancements are also discussed.
}\maketitle

\section{Introduction}

The Liverpool Telescope \citep[LT,][]{2004SPIE.5489..679S} is a 2.0 metre robotic telescope that is operating unattended at the Observatorio del Roque de Los Muchachos 
Observatory on La Palma, Spain. Since robotic operations started in April 2004, the LT has produced data for a variety of 
science programmes using software and instruments that were designed and developed in-house at the Astrophysics Research Institute of LJMU. Achieving first light 
in April 2009, the Fibre-fed RObotic Dual-beam Optical Spectrograph \citep[FRODOSpec,][]{2004AN....325..215M} is the successor to the now decommissioned Meaburn 
Spectrograph and has been a common user instrument on the telescope since February 2010.

FRODOSpec is a bench mounted spectrograph with two optical paths, known as arms, that are utilised by separating the incident light around 5750\AA \ into two bandwidths 
using a dichroic beam-splitter. The light down each arm is collimated, dispersed and focused onto CCDs, with the elements of each optical chain separately optimised for
blue and red light. An optical schematic is shown in Figure \textbf{\ref{fig:frodooptics}}.

Two dispersive elements are available for each arm: a conventional diffraction grating and a higher resolution Volume Phase Holographic (VPH) grating.
The VPH is bonded to a prism so that the light is dispersed at the same angle as the grating, requiring no parts other than the pneumatic stage they are mounted
upon to be moved when selecting between them. Resolving power and wavelength ranges for each arm and dispersive element are shown in Table 
\textbf{\ref{tab:capabilities}}.

Light is transmitted from the focal plane of the telescope to the spectrograph by a bundle of one hundred and forty-four optical fibres. At the telescope focal plane, 
the fibres are arranged in a regular pattern (see Figure \textbf{\ref{fig:frodooptics}}) to form a $12\times12$ integral field unit (IFU), with each fibre coupled to a microlens to minimise light losses. 
Each fibre/microlens covers a field of view on sky of $\sim$ $0.83''\times0.83''$, corresponding to a total field of view of approximately $10''\times10''$. At the input of 
the spectrograph, the fibres are rearranged to form a pseudo-slit which acts as the entrance aperture.

\begin{table*}[ht]
  \begin{center}
    \begin{tabular}{l l l l l}
      \hline
      & & & & \\ [-1ex]
      \textbf{Arm / Dispersive Element} & \textbf{Wavelength Start} & \textbf{Wavelength End} & \textbf{Resolution} & \textbf{Dispersion} \\ [1ex]
      & (\AA) & (\AA) & & (\AA/px) \\ [1ex]
      \hline
      & & & & \\
      Red Grating & 5800 & 9400 & 2200 & 1.6\\ 
      Red VPH & 5900 & 8000 & 5300 & 0.8 \\ 
      Blue Grating & 3900 & 5700 & 2600 & 0.8\\ 
      Blue VPH & 3900 & 5100 & 5500 & 0.35 \\ 
      & & & & \\
      \hline
      \\
    \end{tabular}
    \caption{Wavelength ranges, resolving powers and dispersions for the different FRODOSpec arms/dispersive elements.}
  \label{tab:capabilities}
  \end{center}
\end{table*}

Despite the availability of instrument unspecific software packages to reduce data taken using an integral field spectrograph, the constraints of their generic design
typically limit the highest achievable degree of automation to either manual like IRAF \citep{1992ASPC...25..417V} or semi-automatic 
like R3D \citep{2006AN....327..850S}, kungifu \citep{2007NJPh....9..443B} and P3D \citep{2010A&A...515A..35S}. As these packages do not constitute an end-to-end system 
whereby data products can be produced without the need for human interaction, development of a bespoke pipeline to reduce FRODOSpec data was necessary in 
order that the following objectives could be fulfilled:

\begin{itemize}
 \item To autonomously produce a science-ready data product.
 \item To autonomously produce a ``quicklook'' data product, allowing quick data quality assessment.
 \item To have full quality control over the data products produced.
 \item To provide feedback for the automated LT scheduler.
\end{itemize}

This paper details the second version of the pipeline, deployed in May 2011, and is structured as follows. An explanation of the input data and the output 
data products is given in \textbf{\S \ref{sec:overview}}. \textbf{\S \ref{sec:methods}} outlines the computational methods used to process the data with 
error analysis. Key pipeline performance indicators are presented in \textbf{\S \ref{sec:performance}}. Concluding remarks and possible future enhancements 
to the pipeline are discussed in \textbf{\S \ref{sec:conclusions}}.

\section{Overview}\label{sec:overview}
\subsection{Input data}\label{sec:TheInputData}

\begin{figure*}[!htb]
  \begin{center}
  \includegraphics[scale=1.99]{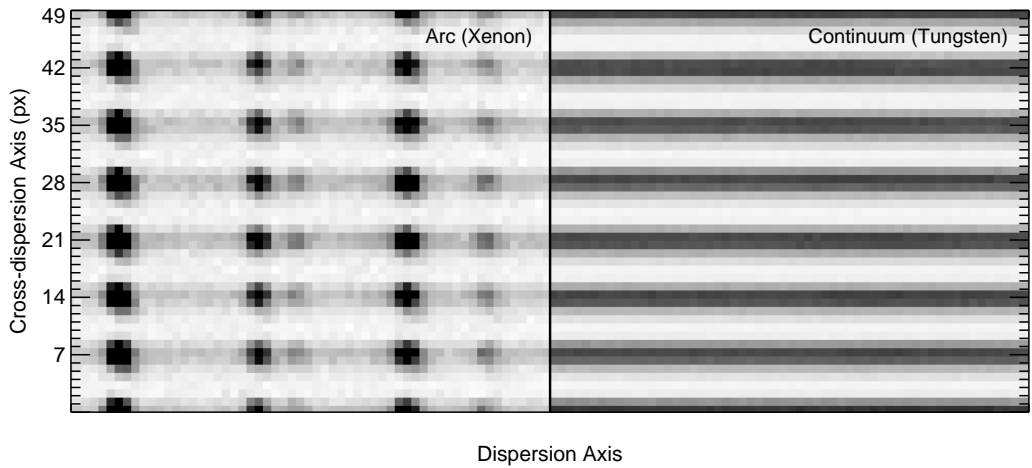}
  \caption{Exposures of Xenon \textbf{(left)} and Tungsten \textbf{(right)} taken using FRODOSpec. The profiles of the fibres can be clearly seen in the 
  Xenon arc frame.}
  \label{fig:frodo_spec}
  \end{center}
\end{figure*}

\begin{figure*}[!htb]
  \begin{center}
  \includegraphics[scale=0.9]{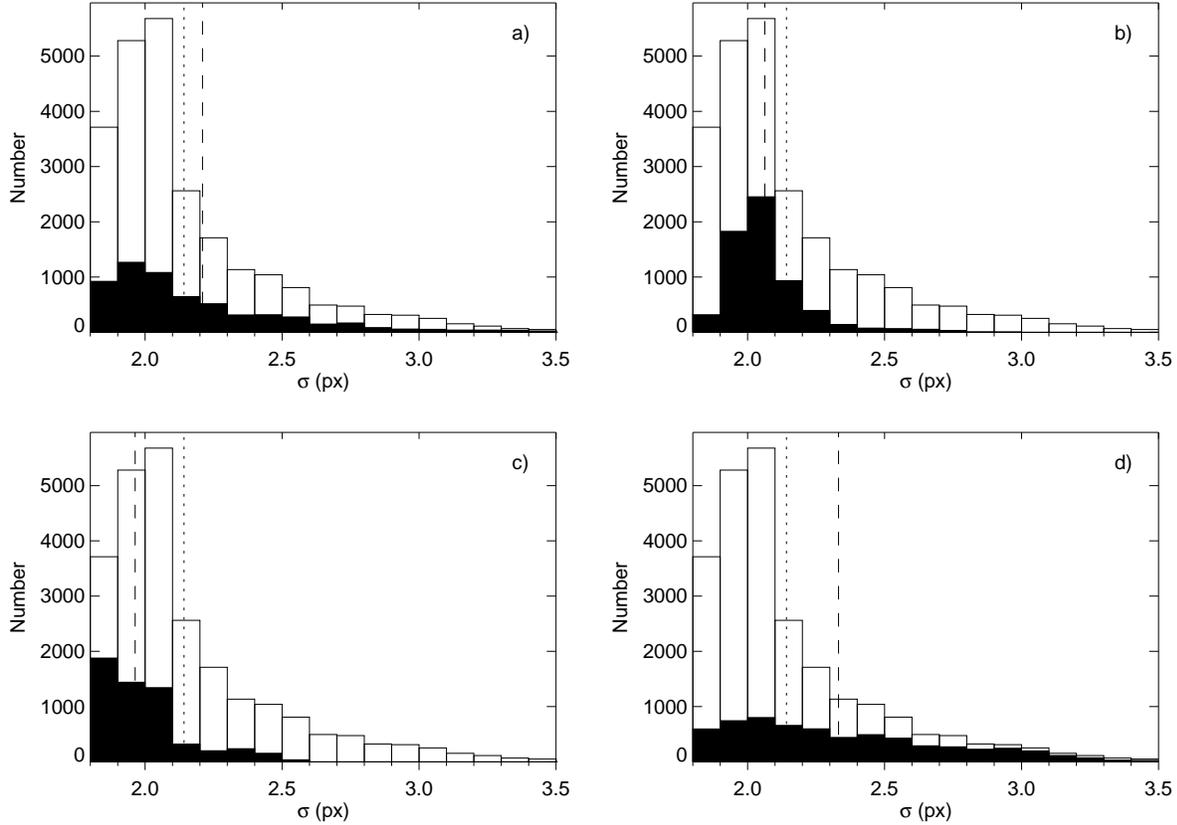}
  \caption{The distribution of the focus measure, $\sigma$. The hollow histograms show the distribution of the entire sample, while the filled histograms show the configuration 
  specific distribution for the: \textbf{a)} red grating, \textbf{b)} red VPH, \textbf{c)} blue grating and \textbf{d)} blue VPH. The 
  dotted lines represent the average for the whole sample (2.14 pixels), while the dashed lines show the averages for the different dispersive elements and arms. 
  (2.21, 2.06, 1.96, 2.33 pixels respectively). The broader form of the blue VPH distribution is indicative that the focus is 
  non-uniform along at least one axis.}
  \label{fig:fibre_fwhm}
  \end{center}
\end{figure*}

\begin{figure*}[!htb]
  \begin{center}
  \includegraphics[scale=0.9]{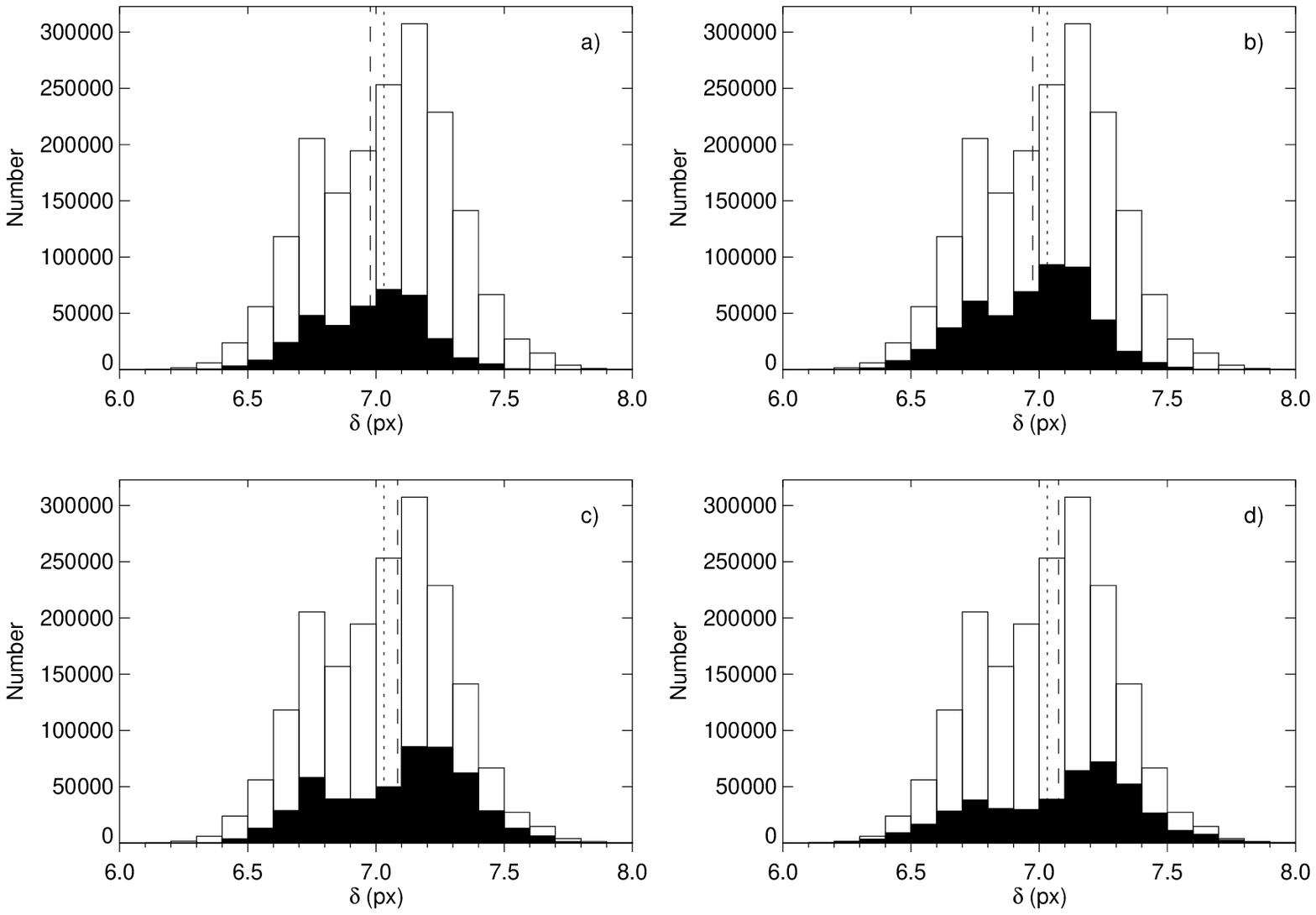}
  \caption{The distribution of the spectral pitch measure, $\delta$. The hollow histograms show the distribution of the entire sample, while the filled histograms show the 
  configuration specific distribution for the: \textbf{a)} red grating, \textbf{b)} red VPH, \textbf{c)} blue grating and \textbf{d)} blue VPH. The 
  dotted lines represent the average for the whole sample (7.03 pixels), while the dashed lines show the averages for the different dispersive elements and arms 
  (6.98, 6.97, 7.08, 7.08 pixels respectively). The bimodal form of the plots is a natural consequence of the misalignment of fibres in the pseudo-slit. If two 
  fibres exhibit a smaller than average inter-fibre distance, there should generally also exist a larger than average inter-fibre distance depending on the position 
  of the two fibres in the pseudo-slit.}
  \label{fig:fibre_sep}
  \end{center}
\end{figure*}

Data taken using an integral field spectrograph (see Figure \textbf{\ref{fig:frodo_spec}}) has two major differences to that of a traditional long-slit:

\begin{enumerate}

\item \textbf{The flux propagates spatially as a function of fibre profile}. In order to describe the spatial distribution of flux, a measure of 
spatial fibre profile width, $\sigma$, and fibre separation, $\delta$, must be introduced. In the context of numerical analysis, the following quantities are 
therefore defined:

\begin{description}
 \item $\sigma$ is the FWHM of the gaussian that best describes the spatial profile of a fibre. Although not perfect, a single gaussian profile is typically 
  a good approximation of the spatial flux distribution (see Figure \textbf{\ref{fig:plot_profile_ct}}), with residual flux accounting for no more than 2\% 
  of the total flux of the distribution. As the width of the fibre profile varies with spectrograph focus, and the focus is dependent on the ambient temperature
  in proximity to the instrument, there is no unique value for $\sigma$. 

  \begin{figure}[!htb]
    \begin{center}
    \includegraphics{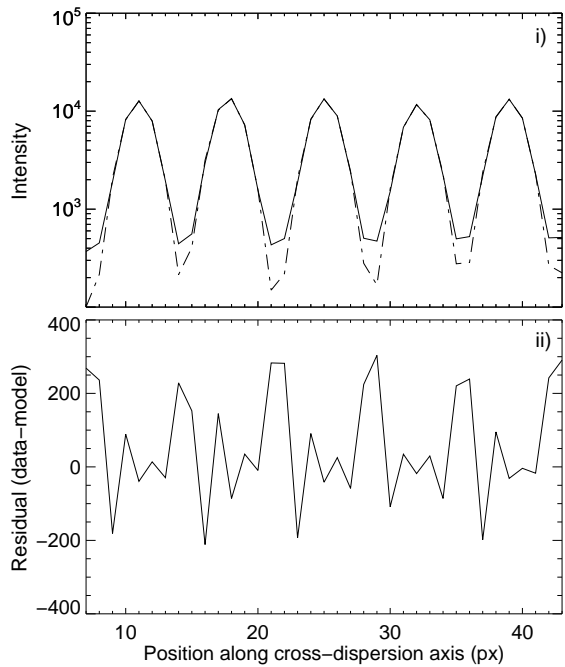}
    \caption{The spatial flux distribution modelled using single gaussians to represent the fibre profiles. The solid and dashed lines in i)
    represent the data and model respectively. The residual flux in the wings of profile cannot be accounted for by a single gaussian, but accounts for no more 
    than 2\% of the total flux distribution.}
    \label{fig:plot_profile_ct}
    \end{center}
  \end{figure}

 \item $\delta$ is the spatial distance between adjacent fibre profile centroids, known as the spectral pitch. Due to fibre positioning errors within the pseudo-slit, 
  there is also no unique value of $\delta$.
\end{description}

To determine the distributions of $\sigma$ and $\delta$, both quantities were measured daily over a three week period for each dispersive element and arm 
(see Figures \textbf{\ref{fig:fibre_fwhm}} and \textbf{\ref{fig:fibre_sep}}). The Starlink package, Figaro, was used to measure $\sigma$, with fibre profile centroids 
calculated by the pipeline (see \S \ref{ss:frfind}) used to measure $\delta$. The optimum focal positions of the CCDs were maintained weekly by remotely driving the
electronic translation stages upon which they are mounted, limiting the timescale over which the focus was allowed to drift. 

\item \textbf{The flux from the slit is spatially incoherent}. As the two-dimensional fibre matrix at the IFU input must be rearranged into a one-dimensional 
slit at the output, the flux from adjacent fibres may originate from different positions on the focal plane.

\end{enumerate}

A complication arising from the combination of these two differences is fibre cross-talk (see \S \ref{sec:crosstalk}).

\subsection{Output data products}\label{sec:TheOutputDataProduct}

Data taken by FRODOSpec is reduced by two sequentially invoked pipelines. The first pipeline, known as the L1, is a CCD processing pipeline that performs bias 
subtraction, overscan trimming and CCD flat fielding. This paper focuses on the second pipeline, known as the L2, which performs the processes specific to the
reduction of data taken using an integral field spectrograph.

The science-ready data product is an eight part multi-extension FITS \citep{2001A&A...376..359H} file with each extension containing a snapshot of the data 
taken at key stages of the reduction process. The lowest tier of reduction product available to the user is the L1 image. The output format is shown in 
Table \textbf{\ref{tab:hdustruct}}.

In addition to the science-ready data product, a composite raster image of the L1\_IMAGE, SPEC\_NONSS and COLCUBE\_NONSS extensions is made available through the 
LT archive website\footnote{\url{http://lt-archive.astro.livjm.ac.uk}}. An example is shown in Figure \textbf{\ref{fig:preview}}.

\begin{table*}[ht]
 \begin{center}
  \begin{tabular}{c l l c c c}
   \hline
     & & & & & \\ [-1ex]
     \textbf{HDU Index} & \textbf{EXTNAME} & \textbf{Format} & \textbf{Wavelength} & \textbf{Throughput} & \textbf{Sky Subtracted?} \\ 
     & & & \textbf{Calibrated?} & \textbf{Corrected?} & \\ [1ex]
     \hline
     & & & & & \\ [-1ex]
   0 & L1\_IMAGE & Image & & & \\ [0ex]
   1 & RSS\_NONSS & RSS & \checkmark & \checkmark & \\ [0ex]
   2 & CUBE\_NONSS & Datacube & \checkmark & \checkmark & \\ [0ex]
   3 & RSS\_SS & RSS & \checkmark & \checkmark & \checkmark \\ [0ex]
   4 & CUBE\_SS  & Datacube & \checkmark & \checkmark & \checkmark \\ [0ex]
   5 & SPEC\_NONSS  & Spectrum & \checkmark & \checkmark & \\ [0ex]
   6 & SPEC\_SS  & Spectrum & \checkmark & \checkmark & \checkmark \\ [0ex]
   7 & COLCUBE\_NONSS  & Image & & \checkmark & \\ [1ex]
   \hline
   \\
   \end{tabular}
   \caption{The format of the science-ready data product. Extensions can be accessed through either their HDU index or corresponding EXTNAME key. 
    Row Stacked Spectra (RSS) frames are used to display each extracted spectrum as a single row of height one pixel. Datacubes reimage the focal 
    plane at each wavelength using the IFU input to output head mappings (two spatial axes, x and y, and the dispersion axis, z). Datacubes are a standard 
    reduction product for integral field spectroscopy (IFS) and many software reduction packages have tools to visualise and manipulate them. The final extension is an 
    IFU fibre matrix image of the focal plane, which is the CUBE\_NONSS extension with a collapsed dispersion axis. If sky subtraction (SS) is unsuccessful, 
    the corresponding HDUs (3,4 and 6) will be blank. Wavelength is calibrated in units of \AA.}
 \label{tab:hdustruct}
 \end{center}
\end{table*}

\subsection{Coding platform}

The pipeline has a command-line interface (CLI), consisting of a series of programs written in C using the GNU Scientific Library \citep[GSL,][]{GSL} and CFITSIO 
\citep{1999ASPC..172..487P} library. 

The compiled binaries are linked through scripts written in TCSH, with the reduction image preview scripts requiring a combination of ImageMagick and GNUplot.

\section{Reduction Method}\label{sec:methods}

L2 reduction largely follows the process that has been previously discussed in the development of both instrument specific 
software packages for spectrographs like VIMOS \citep{2005PASP..117.1271Z}, as well as unspecific ones like R3D, kungifu and P3D.

Specifically, it consists of \textbf{i)} finding and tracing the positions of the fibres at points along the 
dispersion axis, \textbf{ii)} standard aperture extraction, \textbf{iii)} wavelength calibration, \textbf{iv)} fibre throughput correction, \textbf{v)} spectral 
rebinning to a linear wavelength calibration, \textbf{vi)} identification and subtraction of sky-only fibres (only possible if they are available in the field) and 
\textbf{vii)} reformatting of data to desired output format (see \S \ref{sec:TheOutputDataProduct}). A flow chart illustrating the reduction process is shown in 
Figure \textbf{\ref{fig:FrodoL2}}.

No attempt has been made to either remove the contamination from scattered light and cosmic rays, or to correct for differential atmospheric refraction (DAR). A 
synopsis of each of these is given below.

Scattered light is the general term given to light scattered off component optical surfaces of an instrument that has not followed the desired optical path, and can 
be identified in data from a fibre-fed spectrograph as a smooth background between the spatial profiles of the fibres. Removal of scattered light can only be reliably 
achieved if the spectral pitch is sufficiently large that a clean background can be interpolated spatially between the fibre profiles. Although the flux contribution 
from scattered light is dependent on fibre illumination, with larger values for longer exposure times and brighter targets, it has been estimated to contribute no more 
than 2\% of the total flux for any given FRODOSpec observation (see \S \ref{sec:TheInputData}).

Automated removal of cosmic rays from spectrographic data is too unreliable to be implemented using current methods such as L.A.Cosmic \citep{2001PASP..113.1420V} and 
the DCR routine \citep{2004PASP..116..148P}. The accuracy of the removal is dependent on the location of the cosmic ray, with those lying close to strong emission lines 
particularly hard to distinguish and remove cleanly without the iterative manual process of program parameter tweaking and inspection of results.

The extent to which DAR affects an observation can be assessed by measuring the shift in target position at the focal plane for different wavelengths, and is worse for larger 
zenith angles \citep{1982PASP...94..715F} and observations where the target is required to be tracked by the telescope for long periods. DAR must be corrected for by the 
user if spectrophotometric accuracy is required. A method for its correction in IFS has been documented \citep{1999A&AS..136..189A}.

The L2 requires three frames to proceed: a Xenon arc frame, a continuum frame (typically an exposure of a Tungsten lamp) and the target, or science, frame.

\subsection{Constructing the tramline map}\label{sec:traces}

As the true cross-dispersion axis, $x$, and dispersion axis, $\lambda$, are only approximately aligned with the corresponding pixel axes of the CCD, it is required that 
a trace, or tramline map, be made to characterise how the flux from each fibre propagates along both axes of the CCD. This is done by determining the relationship 
between CCD pixel coordinates and the peak location of each fibre profile taken at intervals along the dispersion axis. 

As a target frame will typically have insufficient signal through each fibre over the wavelength range required, an exposure of a continuum lamp is used.
To minimise the effects of shifting spectral positions on these traces due to temperature changes, continuum flats used by the L2 are taken nightly using a 
Tungsten lamp.

\subsubsection{Finding the fibre profile peaks\newline\hspace*{30pt}(frfind)}
\label{ss:frfind}

The first of the pipeline procedures, \textbf{frfind}, is a simple peak finding routine that reads the CCD output row by row along the dispersion 
axis\footnote{The literal distinction between the true dispersion/cross-dispersion and actual CCD axes is ignored in the following discussion.} considering each pixel on the 
cross-dispersion axis in turn and flagging a peak only if the following criteria are all satisfied \citep[cf.][]{2006AN....327..850S}:

\begin{enumerate}
\setlength{\itemsep}{15pt}
\item A specified aperture either side of and spatially contiguous to the current pixel being considered all have fewer counts.
\\\\
Let \textit{I}($i,j$) be the intensity of the currently considered pixel $i$ along the dispersion axis, and $j$ the pixel along the cross-dispersion axis 
locating the peak number $k$ such that determination of the $k^{th}$ peak verifies: 
\\\\
$\textit{I}(i,j) > \textit{I}(i,j+1)$\newline
$\textit{I}(i,j) > \textit{I}(i,j+2)$\newline
\hspace*{29pt}\vdots\newline
$\textit{I}(i,j) > \textit{I}(i,j+(n-1)/2)$\hspace{10pt}\newline\newline
and\newline\newline
$\textit{I}(i,j) > \textit{I}(i,j-1)$\newline
$\textit{I}(i,j) > \textit{I}(i,j-2)$\newline
\hspace*{29pt}\vdots\newline
$\textit{I}(i,j) > \textit{I}(i,j-(n-1)/2)$\newline

where \textit{$n$} is the aperture size in pixels.

A sensible restriction on \textit{$n$} exists, in that it should be less than the spectral pitch. Additionally, as the aperture size is defined by a whole number 
of pixels, the L2 requires that \textit{$n$} be an odd integer.

\item The pixel distance to the previous peak is greater than a pre-specified minimum distance. This criterion is omitted when the first peak in the row is 
being considered.
\\\\   
$j_{k} - j_{k-1} > \mathrm{minimum\;distance}$
\\\\
In order that all fibres can be identified, the value of the minimum distance is set less than the minimum inter-fibre distance.

\item The currently considered pixel value is greater than the value at a pre-specified pixel distance either side by a minimum amount. The optimum value of this amount
is determined automatically by the routine by cycling between pre-specified limits until the maximum number of rows with the correct number of peaks, equal to
the number of fibres, is found.
\\\\
$\textit{I}(i,j) - \textit{I}(i,j-l) > \mathrm{minimum\;amount}$\hspace{10pt}and\newline
$\textit{I}(i,j) - \textit{I}(i,j+l) > \mathrm{minimum\;amount}$
\\\\
where \textit{$l$} is the pre-specified cross-dispersion pixel distance to be checked either side of a candidate peak.

\end{enumerate}
\vspace*{3pt}
Using the GSL least-squares polynomial fitting algorithm, sub-pixel peak locations are determined by finding the position of the maximum,
$j_{k}^{ctd}$, of a parabola fitted to the flux of three pixels at locations along the cross-dispersion axis of $j_{k}-1$, $j_k$ and $j_{k}+1$.

The routine, and consequently reduction, is aborted if the number of rows with the correct number of peaks is less than a pre-specified value.

\subsubsection{Cleaning erroneous entries from the peak list \newline\hspace*{30pt}(frclean)}

The second routine, \textbf{frclean}, is used to remove rows with an appreciable likelihood of containing peaks that have either been incorrectly classified or have 
poorly determined locations. These rows are identified by cycling through the peaks for each row in the peak list generated by \textbf{frfind}, and calculating 
the pixel distance between the cross-dispersion location of the currently considered peak and the average cross-dispersion location for that peak number, 
$\overline{j_{k}^{ctd}}$, determined using all rows. Any peak in a particular row that has a distance larger than a pre-specified maximum distance will restrict all 
peaks in that row from further use by the pipeline.
\\\\
$| j_{k}^{ctd} - \overline{j_{k}^{ctd}} | < \mathrm{maximum\;distance}$
\\\\
The maximum difference parameter is set to allow for the inherent spectral curvature and rotation,

\subsubsection{Fitting polynomials to the fibre profile peaks\newline\hspace*{30pt}(frtrace)}

\begin{figure*}[ht]
  \begin{center}
  \includegraphics[scale=0.95]{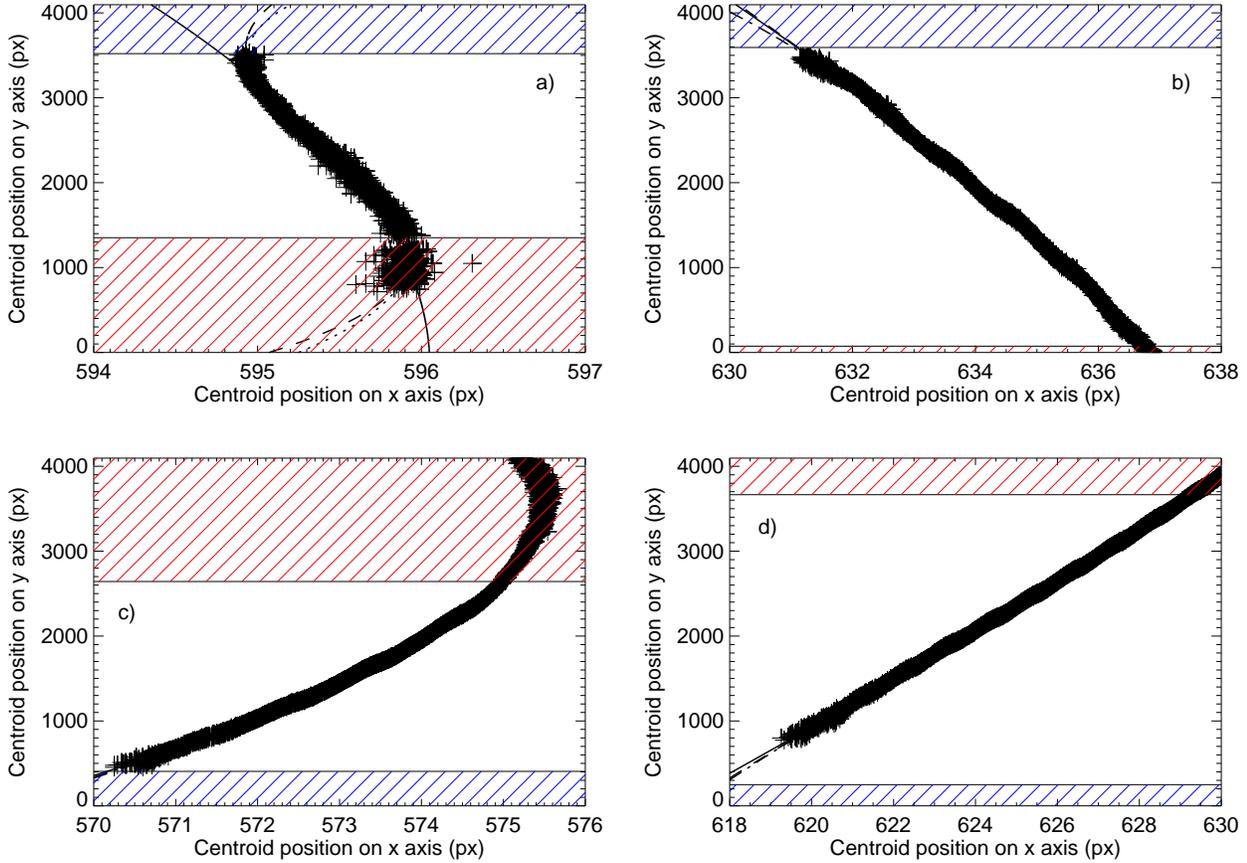}
  \caption{Example polynomial traces using fibre number 78 for the \textbf{a)} red grating, \textbf{b)} red VPH, \textbf{c)} blue grating
  and \textbf{d)} blue VPH configurations. The blue and red hatched regions represent areas on the CCD for which the pixels have a corresponding
  wavelength that is outside of the final wavelength calibration. The solid line represents a fit to the data of order 2, a dashed line of order 3
  and a dotted line of order 4. For the wavelength ranges required, a polynomial order of 2 was found to be suitable for all 
  configurations but the red grating, where an order of 3 was found to better account for the curvature towards the blue end of CCD. Using these orders, the root mean
  square (RMS) discrepancies between the fit and data used in the final wavelength calibration were 0.038, 0.053, 0.047 and 0.053 pixels respectively.}
  \label{fig:plot_traces}
  \end{center}
\end{figure*}

The remaining peak locations in the peak list generated by \textbf{frclean} are binned along the dispersion axis for each fibre. The average cross-dispersion 
locations of the peaks contained within each bin are then calculated and polynomial fits to these positions and the bin centroids are made using the GSL 
least-squares polynomial fitting algorithm. Figure \textbf{\ref{fig:plot_traces}} illustrates examples of these fits
for fibre number 78, central within the pseudo-slit and representative of the other fibres in the bundle.

The quality of the fit depends largely on the number of bins, or bin width, specified. A smaller bin width increases the likelihood of empty bins, where no 
peaks can be found between the bin limits. Conversely, a larger bin width reduces the number of coordinates used in the fitting process. As the curvature is well 
defined by low order polynomial fits (see Figure \textbf{\ref{fig:plot_traces}}), larger bin widths are most suitable for this routine.

\subsection{Extraction of flux \newline\hspace*{23pt}(frextract)}\label{sec:ext}

A standard aperture extraction is performed using the traces determined by \textbf{frtrace}. In standard aperture extraction, all pixels within an aperture 
centered on the trace centroid are assumed to have equal statistical weight. Aperture boundaries that lie inside pixels are accounted for by adding the 
fractional flux from the corresponding pixel. 

For a raw image, $C$, the data values, $D$ after processing by the L1 CCD pipeline are given by \citep[cf.][]{1986PASP...98..609H}:

\begin{equation*}
 D_{x\lambda} = \frac{C_{x\lambda}-B_{x\lambda}}{F_{x\lambda}}
\end{equation*}

where $B$ is the master bias image and $F$ is the flat field image. The integrated flux for each spectrum, $f^{s}$, and variance, $\mathrm{var}^{s}$, between the 
spatial aperture limits $x_{1}$ and $x_{2}$ for a wavelength $\lambda$ is:

\begin{equation*}
 f^{s}_{\lambda} = {\sum^{x_{2}}_{x = x_{1}}{D_{x\lambda}}}
\end{equation*}

\begin{equation*}
 \mathrm{var}^{s}_{\lambda} = {\sum^{x_{2}}_{x = x_{1}}{V_{x\lambda}}}
\end{equation*}

with the variance on an individual pixel, $V_{x\lambda}$, approximately given by:

\begin{equation*}
 V_{x\lambda} = r^{2} + \frac{|D_{x\lambda}|}{G}
\end{equation*}

where $r$ is the readout noise (e$^{-}$), assumed to be constant across the CCD, and $G$ is the gain of the CCD (e$^{-}$ / ADU). As a large number of frames 
are stacked to generate both the master bias and flat field frames, their resulting error contribution is assumed to be neglible. 

The L2 uses a 5 pixel extraction aperture, chosen as a comprimise between minimising fibre cross-talk and maximising the amount of flux recovered from 
the fibre profile. 

\subsubsection{Fibre cross-talk}\label{sec:crosstalk}

Cross-talk occurs when the profiles of the fibre overlap spatially. The severity of cross-talk is dependant on a variety of factors including $\sigma$ and $\delta$. 

\begin{figure}[ht]
  \begin{center}
  \includegraphics{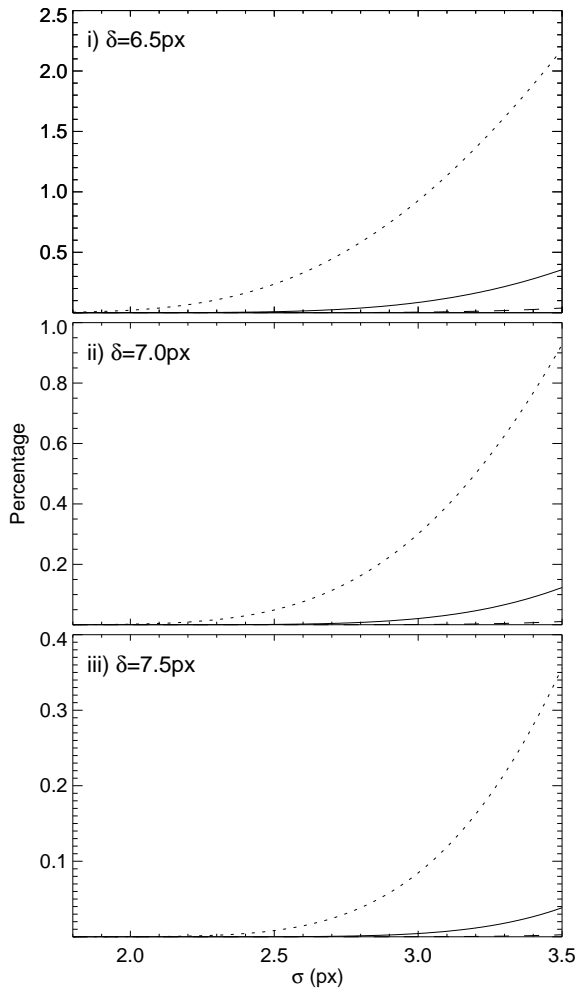}
  \caption{Percentage of total flux contained in an adjacent aperture for varying $\sigma$, $\delta$ and $\gamma$. The dotted, solid and dashed 
  lines represent a 7 pixel, 5 pixel and 3 pixel extraction aperture respectively.}
  \label{fig:crosstalk}
  \end{center}
\end{figure}

\begin{figure}[ht]
  \begin{center}
  \includegraphics{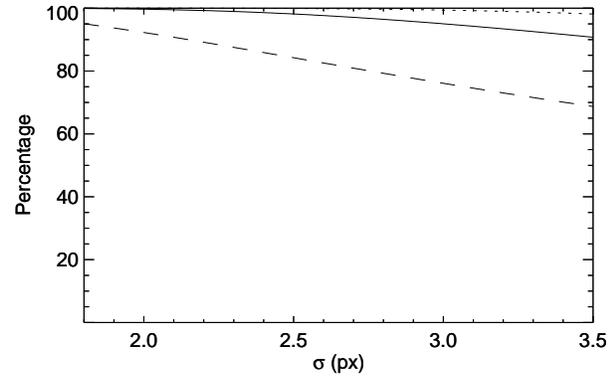}
  \caption{Percentage of total flux recovered in the aperture for varying $\sigma$ and $\gamma$. The dotted, solid and dashed 
  lines represent a 7 pixel, 5 pixel and 3 pixel extraction aperture respectively.}
  \label{fig:fluxrecovered}
  \end{center}
\end{figure}

To quantify this effect for FRODOSpec, an approximate two aperture analysis is used. Each aperture has a width in pixels
of $\gamma$, with the centre of the apertures separated by the spectral pitch, $\delta$. A fibre is placed with centroid at $x=\delta$ and the corresponding flux, 
$f_{CT}$, recovered by an aperture centered at $x=0$ is calculated. As previous, the fibre profile is modelled using a single gaussian such that:

\begin{equation*}
  f_{CT} = {\int^{\gamma/2}_{-\gamma/2} e^{\dfrac{-(x-\delta)^2}{2\sigma^2}}\,dx}
\end{equation*}

which is then represented as a percentage of the total flux. Figure \textbf{\ref{fig:crosstalk}} shows how this cross-talk quantifier varies
for different $\delta$, $\sigma$ and $\gamma$. Figure \textbf{\ref{fig:fluxrecovered}} shows how the flux recovered by an aperture centered on the fibre 
centroid varies for different $\delta$ and $\gamma$.

To assess if the effect of cross-talk must be taken into account, two cases representing the worst ($\sigma$ = 3.5px, $\delta$ = 6.5px) and average 
($\sigma$ = 2.14px, $\delta$ = 7.03px) data conditions are presented for a 5 pixel extraction aperture.

In the worst case, $\sim$90\% of the flux is recovered with a maximum of 0.4\% of the flux contained singly within adjacent apertures. In the average case, 
more than 99\% of the flux is recovered with less than 0.1\% of the flux contained singly within adjacent apertures. 

In order to justifiably apply these figures to real data, an additional consideration must first be made regarding the ability of the pipeline to
accurately locate and trace the position of each fibre within the data. For this to be estimated, it is required that the true peak locations be known a 
priori.

Information regarding the spectral flux distribution, the order of tracing polynomial, the spectral pitches and the spectral and spatial FWHM distributions was used to 
generate simulated data for each configuration. This information was determined using the L2 intermediate reduction products.

\begin{figure*}[ht]
  \begin{center}
  \includegraphics[scale=0.95]{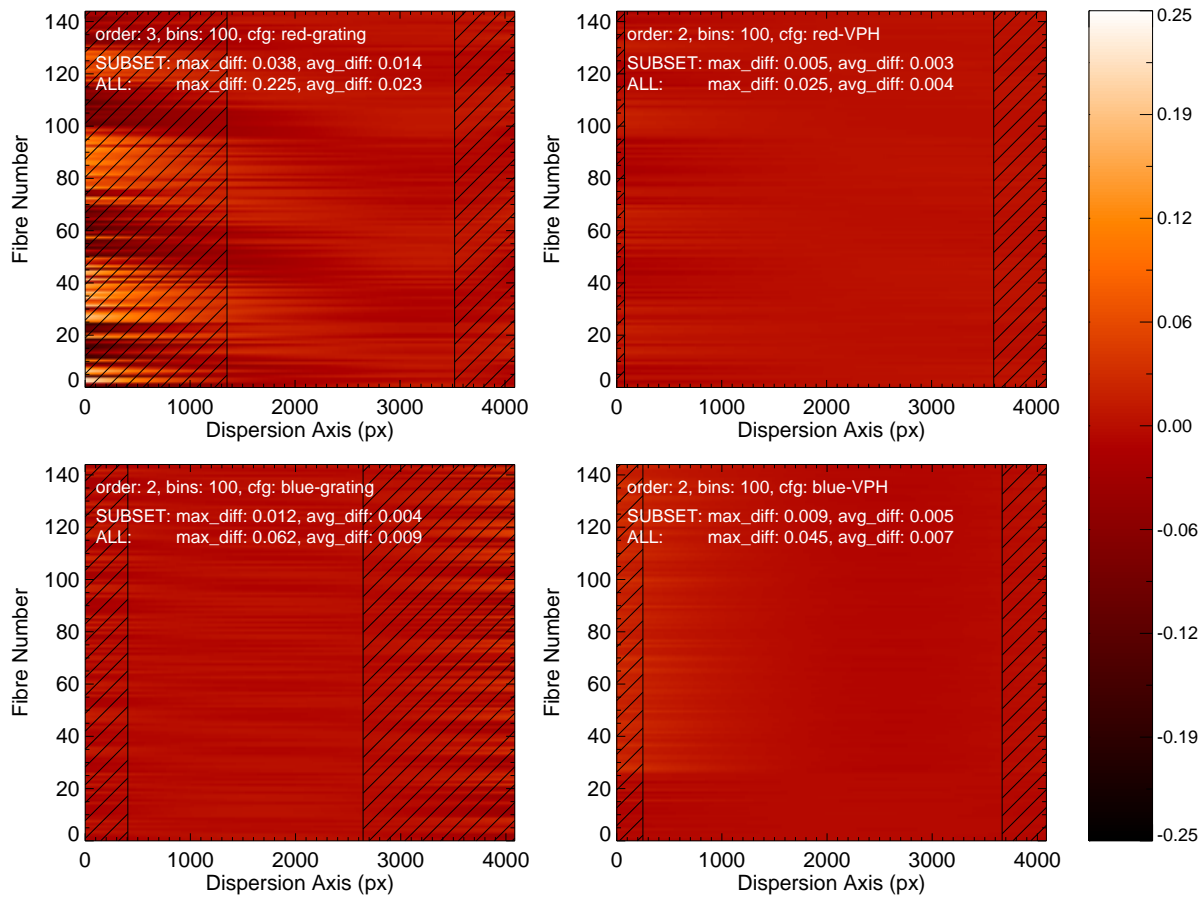}
  \caption{Centroid difference images using simulated data for each configuration. Differences were calculated by subtracting the known peak locations from the
  centroids determined using the polynomial tracing coefficients, and are therefore representative of the conjoint error in centroiding (\textbf{frfind}) and tracing 
  (\textbf{frtrace}). The hatched regions represent areas on the CCD for which the pixels have a corresponding wavelength that is outside of the final wavelength 
  calibration. Maximum and average difference values within these calibrations (SUBSET) and for the entire range of the CCD (ALL) are shown for each figure. As expected, 
  the largest discrepancy is present in the red grating data where a significant fraction of the CCD, the majority of which lies outside of the final calibration, 
  is insufficiently well illuminated to determine an accurate centroid.}
  \label{fig:sim_traces}
  \end{center}
\end{figure*}

After processing the simulated data using the pipeline, the maximum and average difference values were calculated by subtracting the known peak locations from the 
locations determined using the tracing coefficients. The results are shown in Figure \textbf{\ref{fig:sim_traces}}. 

As the discrepancies within the final wavelength calibrations are small, centroiding errors introduced by locating (\textbf{frfind}) and tracing ({\textbf{frtrace}) the 
fibres can be ignored and the previous flux extraction and spatial cross-talk estimations are considered accurate. Consequently, cross-talk is considered a negligible
effect for FRODOSpec and hereafter not considered.

\subsection{Arc fitting \newline\hspace*{23pt}(frarcfit)}\label{sec:frarcfit}

The L2 automatically attempts to fit dispersion solutions.  

Spectra in the arc RSS frame are first cross-correlated to remove zeroth order offsets arising from both optical distortions and small errors in
alignment of the fibres within the slit. This removes the gross curvature and aligns the spectra to $\pm1$px. The cross-correlation is performed using a limited window 
size, decreasing execution times and reducing the influence of bad data on the determination of the offset. An example of this process is shown in panels \textbf{i)} 
and \textbf{ii)} of Figure \textbf{\ref{fig:plot_arc_accuracy}}. The routine then indentifies candidate lines in the data from each spectrum using the criteria outlined 
in \S \ref{ss:frfind} and a few additional constraints described below.

\begin{figure*}[ht]
  \begin{center}
  \includegraphics[scale=0.96]{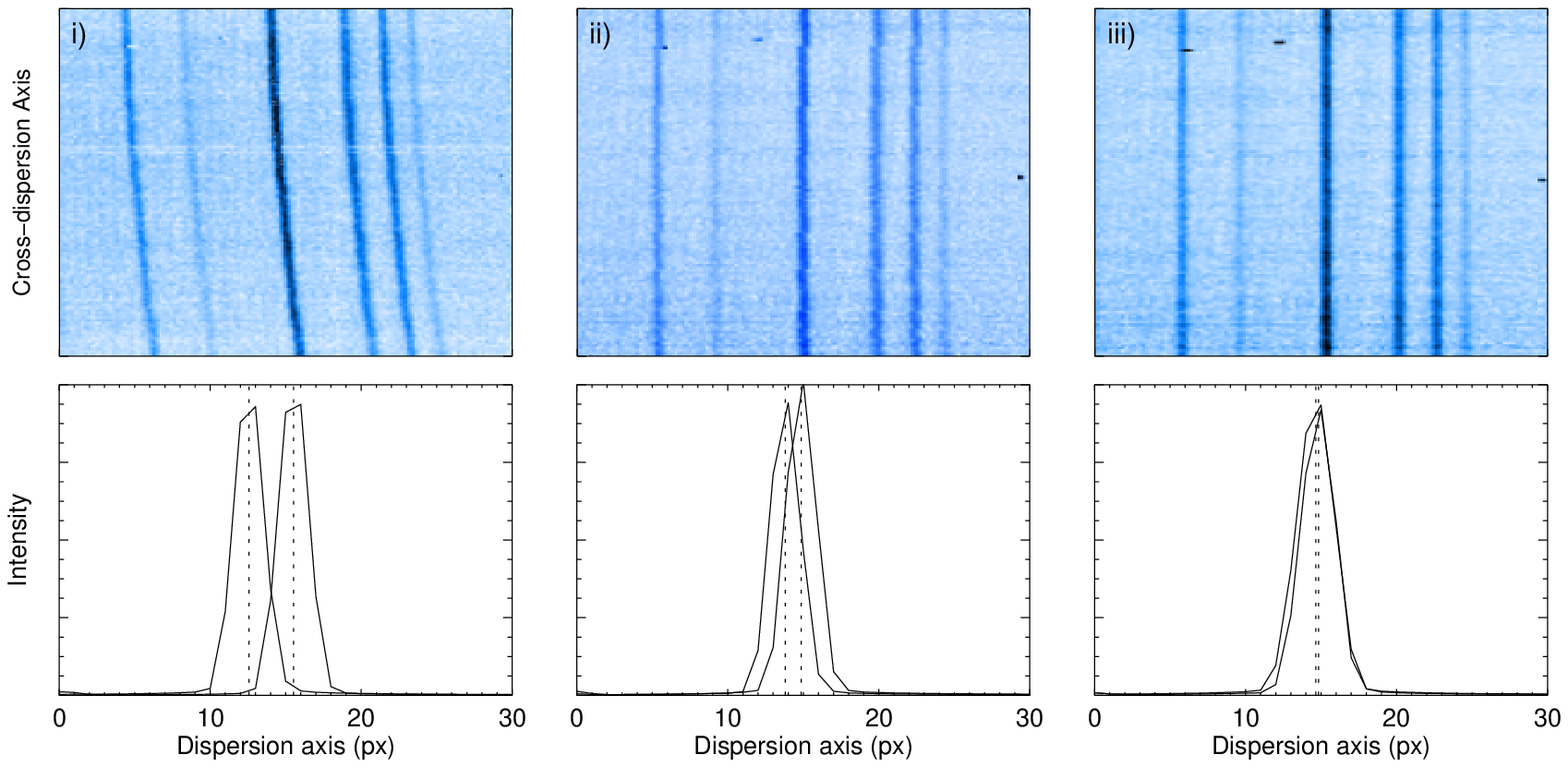}
  \caption{The arc frame at various points in the arc fitting and rebinning process. The top panels show the arc RSS frames, while the bottom panels show the
  maximum difference along the dispersion axis between peaks in the arc RSS frame for a single emission line. \textbf{i)} The gross spectral curvature 
  ($\sim$3px) is evident before the frame is processed but this is removed in \textbf{ii)} by cross-correlating a single spectrum with the others, reducing 
  the maximum difference to $\le$1px. Dispersion solutions are then found for each spectrum in the RSS frame and the flux rebinned to a linear wavelength 
  solution in \textbf{iii)}. Post-rebin maximum differences for the 3 week dataset were calulated for the red grating, red VPH, blue grating and blue VPH and found
  to be 0.23$\pm$0.06, 0.29$\pm$0.31, 0.28$\pm$0.22 and 0.32$\pm$0.44 pixels (or 0.37$\pm$0.1, 0.23$\pm$0.25, 0.22$\pm$0.18, 0.11$\pm$0.15 \AA) respectively.}
  \label{fig:plot_arc_accuracy}
  \end{center}
\end{figure*} 

A candidate line is defined as a cross-spectrum set of spatially contiguous peaks. Spatial contiguity is determined by considering each peak in the distribution of a 
single spectrum of the RSS frame, and checking to see if there exists a peak at the same location along the dispersion axis, within a pre-specified tolerance, for all 
remaining spectra.

Candidate lines are checked against a reference arc line list, which contains information on the wavelength of the emission lines and their 
approximate pixel position. In order for an identified candidate line to be associated with one from the list, the distance between the two must be 
less than that of a pre-specified number of pixels. 

As incorrect line association significantly affects the accuracy of the fits, both the peak-finding criteria and reference arc lines are selected so as to limit the
number of detections and associations made to only those that are deemed most suitable for the purpose of automatic arc fitting. Suitability is determined by 
careful inspection of the data, in order that weak, saturated and crowded lines are not used in this process. It should be noted that a failure to find one of the lines 
is an indicator that the nature of the data has changed significantly since the reference arc line list was made, the most likely cause being a spectral shift due to 
substantial temperature changes (the L2 can tolerate small shifts). 

Once all candidate lines have been identified, two further checks are made to ensure the arc fitting routine produces a reliable and consistent solution:

\begin{enumerate}
 \item The number of lines matched must be greater than a pre-specified number.
 \item The RMS wavelength distance between identified lines and the RMS wavelength distance between lines from the reference arc line list must be within a
 pre-specified amount. 
\end{enumerate}

The joint success of these criteria ensures that the L2 identifies a reasonable quantity and spread of lines taken from within the entire wavelength range. If either 
of these criteria are not met, the reduction is aborted. 

On success, a set of linear equations in a pixel dimension, $x$, and wavelength, $\lambda$, are solved using the GSL least-squares 
polynomial fitting algorithm. Suitable fitting orders were determined by manually finding the dispersion solutions using the arc routine in the Figaro 
package, and selecting an order that minimised the average line residual RMS without introducing too many free parameters (see Figure \textbf{\ref{fig:RMS}}).

\begin{figure}[ht]
  \begin{center}
  \includegraphics{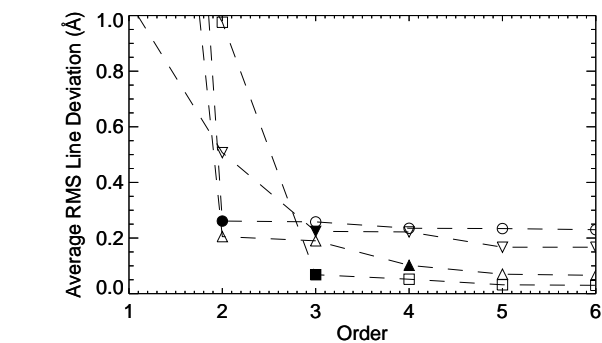}
  \caption{Average order-dependent RMS line deviations for the red grating (circle), red VPH (triangle), blue grating (upside down triangle) and blue VPH (square) 
  configurations. Axes ranges have been selected to highlight the RMS line deviation differences between orders 2 through 6. Filled symbols represent the final orders 
  chosen for each configuration. }
  \label{fig:RMS}
  \end{center}
\end{figure}

\subsection{Throughput correction \newline\hspace*{23pt}(frcorrectthroughput)}
\label{ss:frcorrect}

Fibre defects and misalignments during positioning in the pseudo-slit cause the throughput to vary from fibre to fibre. To correct for this, normalisation coefficients 
are applied to each data value in the target RSS frame. These coefficients are determined by division of the total flux of the median spectrum by the 
total flux of each spectrum in the continuum RSS frame. 

Average normalisation coefficients for the three week dataset are shown in Figure \textbf{\ref{fig:throughputs}}. The percentage relative standard deviation (\%RSD) was 
also calculated using all continuum frames and averaged for all wavelengths along the dispersion axis that were within the post-rebin wavelength calibration 
limits. \%RSD is the absolute value of the coefficient of variation, expressed as a percentage, i.e.
\begin{equation*}
 \%RSD = \dfrac{\sigma}{\bar{x}} \times 100
\end{equation*}

The averaged \%RSDs before and after throughput correction are shown in Table 
\textbf{\ref{tab:throughputs}}. 

\begin{figure*}[ht]
  \begin{center}
  \includegraphics{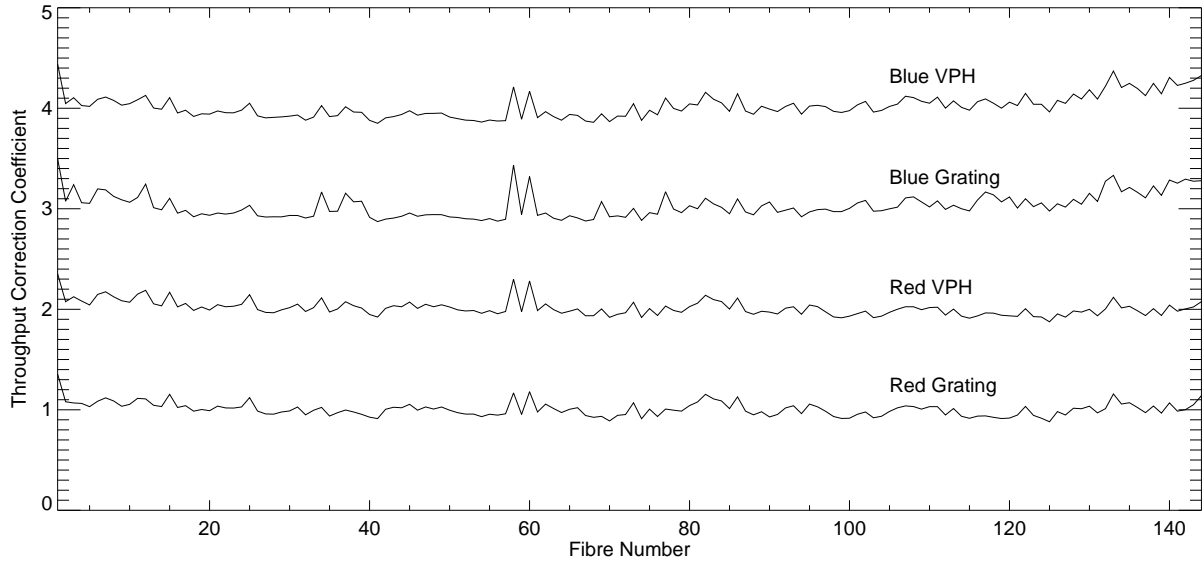}
  \caption{Average throughput normalisation coefficients calculated using Tungsten RSS frames for each arm and dispersive element. The graphs have been offset from each 
  other by unity for clarity. Fibres 58 and 60 are known damaged fibres, and are corrected for by allocation of larger coefficients by the \textbf{frcorrectthroughput}
  routine. The gradual curvature in the blue data may be indicative of vignetting in the optics, with larger throughput coefficients being applied to fibres at the ends
  of the pseudo-slit to compensate.}
  \label{fig:throughputs}
  \end{center}
\end{figure*}

\begin{table}[ht]
 \begin{center}
  \begin{tabular}{l l l}
   \hline
   & & \\ [-1ex]
   \textbf{Arm / Dispersive Element} & \textbf{RSD (Before)} & \textbf{RSD (After)} \\ [0ex]
   & (\%) & (\%) \\ [1ex]
   \hline
   & \\
   Red Grating & 12.3 & 9.7 \\
   Red VPH & 9.6 & 6.2 \\
   Blue Grating & 11.8 & 2.6 \\
   Blue VPH & 12.8 & 5.9 \\
   & \\
   \hline
   \\
   \end{tabular}
   \caption{Average \%RSDs determined using Tungsten RSS frames before and after throughput correction. The routine is less effective at normalising
   the red arm throughputs due to the CCD fringing pattern, which introduces an appreciable spatial dependency to the total fibre fluxes in well illuminated continuum 
   frames.}
 \label{tab:throughputs}
 \end{center}
\end{table}

\subsection{Spectral rebinning \newline\hspace*{23pt}(frrebin)}

In order to add together flux from different spectra, it is first required to apply a single wavelength solution to all the spectra. This is done by rebinning the flux 
along the dispersion axis for each spectrum in the target RSS frame. 

Using configuration specific wavelength start, end and pixel scale parameters (see Table
\textbf{\ref{tab:capabilities}}), the value of the rebinned flux at determined wavelength intervals is calculated 
by linearly interpolating the flux between the two straddling wavelength values. Flux is conserved by calculating the total fluxes in each spectrum before and 
after rebinning, and applying a conservation factor. Example output is shown in the panel \textbf{iii)} of Figure \textbf{\ref{fig:plot_arc_accuracy}}. RMS line 
residuals using a manual and automatic fit are shown in Figure \textbf{\ref{fig:plot_arc_accuracy_multi}}.

\begin{figure*}[ht]
  \begin{center}
  \includegraphics[scale=0.96]{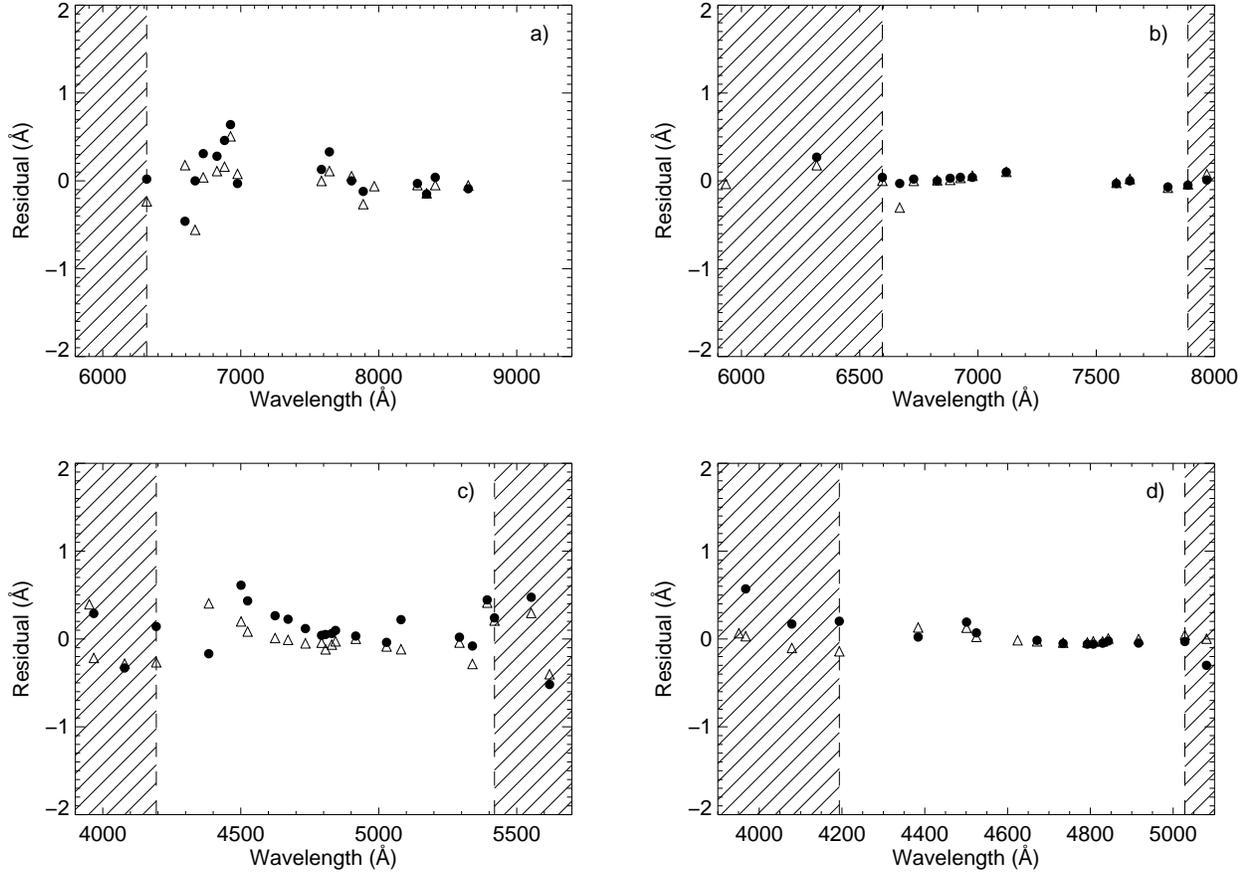}
  \caption{RMS line residuals for all known lines using a manual and automatic fit for the \textbf{a)} red grating, \textbf{b)} red VPH, \textbf{c)} blue grating
  and \textbf{d)} blue VPH configurations. The hatched regions represent areas where the automatic fit is an extrapolation (due to \textbf{frarcfit} only using a
  selection of the known available lines). The hollow triangles represent the residuals of the lines determined using the manual fit, while the filled circles represent 
  the residuals of the lines determined using the automatic fit. The RMS line residuals for the manual fit were 0.26, 0.10, 0.22 and 0.07 \AA \ respectively. 
  The RMS line residuals for the automatic fit were 0.27, 0.08, 0.28 and 0.19 \AA \ respectively. Residuals for lines that could not be accurately centroided
  are omitted from these plots.}
  \label{fig:plot_arc_accuracy_multi}
  \end{center}
\end{figure*}

\subsection{Sky identification and subtraction \newline\hspace*{23pt}(fridsky)}\label{sec:frsubsky}

As FRODOSpec's IFU has a small field of view, a routine that attempts to mask the target and interpolate a sky background would be unreliable as a) the target may be located 
toward the edges of the IFU and b) the target may be extended, making the interpolation ill-defined. The L2 does attempt sky subtraction, but instead uses a sky-only
fibre dataset to calculate the median flux at each wavelength. This contribution is then subtracted off all the spectra in the target RSS frame, leaving the target-only
flux. 

To construct the sky-only fibre dataset, the fluxes contained within each spectrum in the target RSS frame are summed along the dispersion axis to produce a dataset, $X$, containing the 
total fluxes, $I_{T}$, of each fibre. An iterative $\sigma$-clipping algorithm is then used to remove fibres containing target flux from the dataset to create a
sky-only fibre dataset, $X^{i}_{s}$, for each iteration $i$ where $X^{i}_{s} \subseteq X$. For a fibre to qualify as containing target flux within an iteration, its 
total flux must be greater than $n_1$ sigma from the mean of the determined sky-only fibre dataset for the iteration, $\overline{X^{i}_{s}}$: 

\begin{equation*}
  I_{T} > \overline{X^{i}_{s}} + n_1\sigma[X^{i}_{s}]
\end{equation*}

With the first iteration, $X^{1}_{s}$, proceeding as $X^{1}_{s} = X$, the process is repeated using each resulting dataset until no more fibres are identified as containing target flux
and the final iteration of the sky-only dataset, $X^{f}_{s}$, is found. The L2 uses a global detection limit of $n_1=2$.

Aside from a check to ensure that $X^{f}_{s} \neq X$, a comparison between the mean of the final iteration of the sky-only fibre dataset, 
$\overline{X^{f}_{s}}$, and the median of the complete fibre dataset, $\widetilde{X}$, is made to check if the average total sky background flux is statistically 
similar to the total flux of the 77th and 78th brightest fibres:

\begin{equation*}
 \overline{X^{f}_{s}} - n_2\sigma[X^{f}_{s}] < \widetilde{X} < \overline{X^{f}_{s}} + n_2\sigma[X^{f}_{s}]
\end{equation*}

where the value of $n_2$ selected determines to what degree the two values must be statistically similar. It is currently set to 1.

These checks serve as a catch for cases where a) there is no target in the IFU b) the source is extended such that the flux is invariant across the entire IFU
and c) the observation of the target fills more than 50\% of the IFU. Cases a) and b) are computationally identical.

To assess the effectiveness of the routine, exposures of the blank night sky were taken using all configurations. A modified version of the routine was used so
that fibres from half of the IFU were forcibly assigned as sky and half were assigned as target. Bypassing the usual checks, analysis was then carried out on the 
faux target fibre dataset. The results are shown in Figure \textbf{\ref{fig:subsky}}. OI line emission and OH bands \citep{1950ApJ...111..555M} have been marked.

\begin{figure*}[ht]
  \begin{center}
  \includegraphics[scale=0.96]{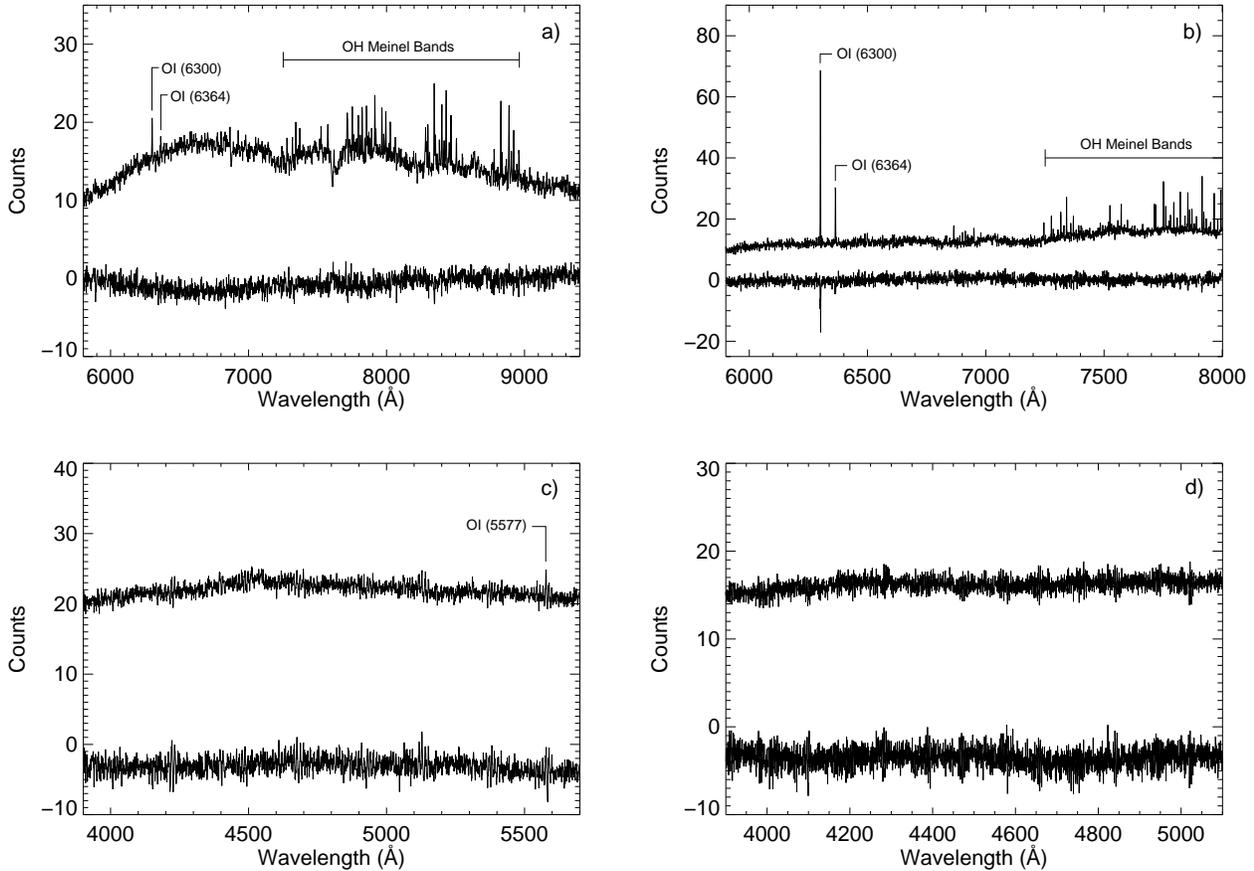}
  \caption{Median non sky subtracted and sky subtracted spectra of the blank night sky for the \textbf{a)} red grating, \textbf{b)} red VPH, \textbf{c)} blue grating
  and \textbf{d)} blue VPH configurations. The flux from strong sky emission lines, such as the OI (6300\AA) line in the spectrum of the red VPH exposure, has been
  oversubtracted. The non-zero offset in the blue configurations is due to scattered light, which introduces a small ($<$1 ADU/600s) non-uniform flux ramp across 
  the CCD.}
  \label{fig:subsky}
  \end{center}
\end{figure*}

\subsection{Formatting of final data product \newline\hspace*{23pt}(frreformat)}\label{sec:frreformat}

Using intermediate non sky subtracted and sky subtracted target RSS frames, the science-ready data product is assembled in accordance with the format shown in Table 
\textbf{\ref{tab:hdustruct}}. In the one-dimensional spectra, only the flux from the top $n$ brightest fibres is summed. To reduce the influence of cosmic rays 
when identifying the brightest fibres, each spectrum in the target RSS frame is individually smoothed along the dispersion axis using a median $10\times1$ pixel 
boxcar filter. The summation is then done using the unsmoothed data. 

For observations of point sources, a smaller value of $n$ reduces the contamination from cosmic rays, as well as generally improving the S/N. For larger values of $n$, 
the total flux recovered is greater. The percentage of the total flux recovered by a subset of fibres is dependent on where the centroid of the target PSF is located on 
the IFU. The two limiting cases are where i) the PSF centroid is located at the centre of a fibre and ii) the PSF centroid is located exactly between four fibres. 
Under the assumption of a gaussian PSF, the percentage of the total flux recovered in i) will generally recover more flux than in ii), except when $n=\{4,5,6\}$.

As is shown in Figure \textbf{\ref{fig:IFU_centroid}}, a summation of the flux from five fibres ($n = 5$) is a good comprimise, as the flux from a point source will be 
contained ($\sim$97.5\% of the total) under average LT seeing conditions of $0.8'' - 1.3''$ regardless of where the centroid is located on the IFU.

\begin{figure}[ht]
  \begin{center}
  \includegraphics{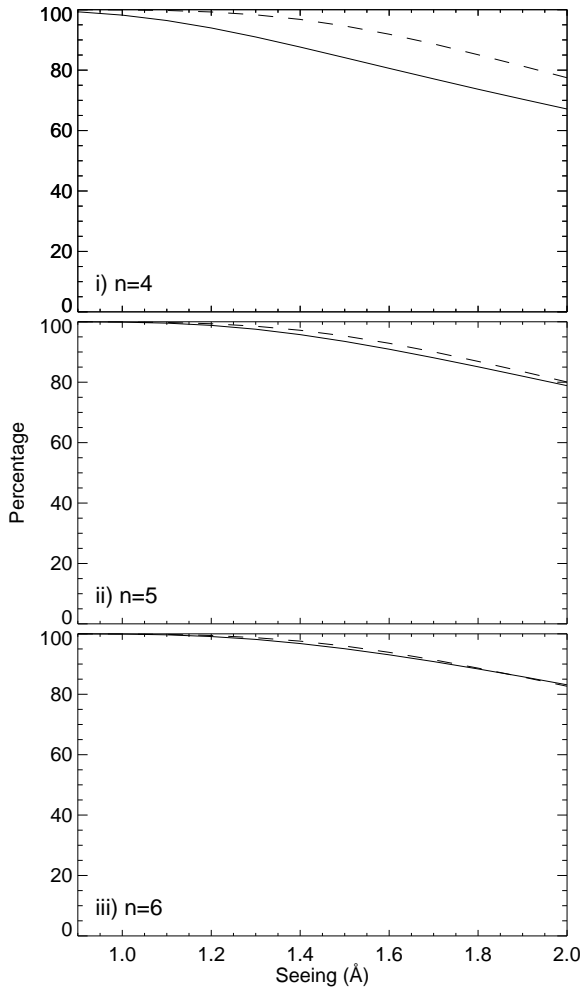}
  \caption{Percentage of the total flux recovered by $n$ fibres for different seeing conditions under the assumption of a gaussian PSF. This quantity is dependent on the 
  location of the PSF centroid on the IFU. The solid line represents the case where the centroid is at the centre of a fibre, and the dashed line represents the case where
  the centroid lies exactly between four fibres.}
  \label{fig:IFU_centroid}
  \end{center}
\end{figure}

For observations where spatially resolved spectroscopy is required, such as the case for extended sources, the one-dimensional spectra may not be a useful data product. 

\section{Performance}\label{sec:performance}

\subsection{Success rates}

The second version of the L2 is now fully integrated into the daily automated FRODOSpec reduction process. Two types of
failure have been seen to occur, with the following important distinction between them:

\begin{itemize}
\item A partial failure is deemed as a failure that still produces an L2 output file. By far the most common partial failure to 
occur so far is unsuccessful sky subtraction. It is worth noting that although this may be considered a failure in the
reduction process, it is by no means the result of an error in the pipeline and is actually the result of constraints
placed in the sky subtraction routine (\textbf{frsubsky}) to ensure that it only occurs when considered statistically 
sensible (see \S \ref{sec:frsubsky}).

\item A critical failure is deemed as a failure that halts the production of L2 output data products. Such a failure may result from the inability of the pipeline
to match enough candidate lines to lines in the corresponding reference arc line list (see \S \textbf{\ref{sec:frarcfit}}).
\end{itemize}

All data taken between 17/06/2011 and 01/08/2011 has been processed using the L2, allowing success rates to be determined for this period. 
Of the 564 frames reduced, 4 experienced a critical failure ($\sim$0.7\% failure rate) whilst 205 experienced a partial failure 
($\sim$36.3\% failure rate).

It is prudent at this point to reiterate that a partial failure resulting from unsuccessful sky subtraction does not impede the processing of the output data products. 
As such, the more accurate indicator of reduction success is the success rate of the pipeline subject to critical failure only. That is to 
say, 99.3\% of observations taken with FRODOSpec will be reduced by the L2 to an extracted, throughput corrected and wavelength calibrated spectrum.

\subsection{Execution times}

The speed at which the pipeline executes is reliant on the specification of the system under which it is being executed and the load the processor is experiencing. L2 reduction occurs
on the LT Archive machine, which has four Intel Xeon 3.20GHz processors and 4GB of RAM. Shown in Table \textbf{\ref{tab:executiontimes}} are the average 
execution times for this machine taken from the 3 week dataset.

\begin{table}[ht]
 \begin{center}
  \begin{tabular}{l l l}
   \hline
   & & \\ [-1ex]
   \textbf{Arm / Dispersive Element} & \textbf{Time} & \textbf{Frames} \\ [0ex]
   & (seconds) & \\ [1ex]
   \hline
   & \\
   Red Grating & 32.6 & 61 \\
   Red VPH & 32.2 & 235 \\
   Blue Grating & 37.4 & 60 \\
   Blue VPH & 36.4 & 208 \\
   & \\
   \hline
   \\
   \end{tabular}
   \caption{Average FRODOSpec execution times for each dispersive element and arm.}
 \label{tab:executiontimes}
 \end{center}
\end{table}

\section{Conclusions}\label{sec:conclusions}

This paper has explained the characteristics of data taken with FRODOSpec, and has presented a brief overview of the computational methodology used to develop a fully 
autonomous pipeline to reduce it. Throughout the paper, representative data taken using each dispersive element and arm was used to assess the 
magnitude of errors and how they propagate through the pipeline. The effect of cross-talk has been discussed, and its contribution found to be negligible. The L2 output 
data products have also been described, along with some key performance indicators. 

The L2 is in a state of continued development. Possible future enhancements include the optimal extraction of flux, which has been shown to increase S/N with a 
corresponding maximum increase in effective exposure time of up to 70\% \citep{1986PASP...98..609H}. All enhancements and addenda will be published on the instrument 
website at \footnotesize\burl{http://telescope.livjm.ac.uk/Info/TelInst/Inst/FRODOspec/}\normalsize.

\acknowledgements{RMB acknowledges STFC for a postgraduate studentship.}

\appendix {

  \newpage

  \section{FRODOSpec Optical Schematic}
  \begin{figure*}[ht]
  \begin{center}
  \includegraphics[scale=0.80]{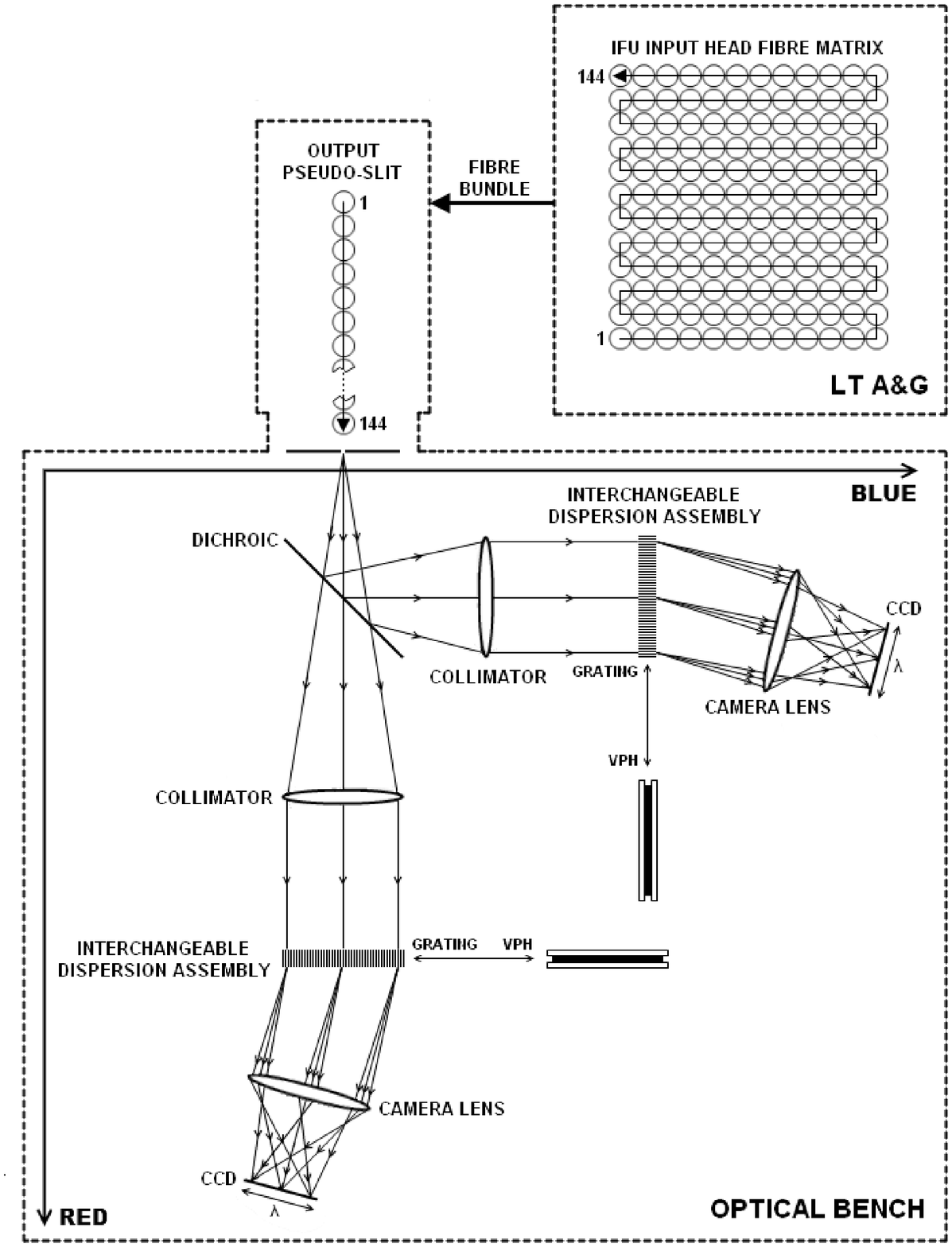}
  \caption{FRODOSpec optical schematic. The FRODOSpec fore-optics are mounted on the telescope acquisition and guidance (A\&G) box. The spectrograph optics are 
  bench mounted and fed from the IFU input head by a bundle of 144 fibres. The pattern used to rearrange the fibres from a two dimensional matrix to a
  one dimensional pseudo-slit is shown.}
  \label{fig:frodooptics}
  \end{center}
  \end{figure*}

  A FRODOSpec optical schematic is shown in Figure \textbf{\ref{fig:frodooptics}}. 

  \section{FRODOSpec L2 Reduction Preview Example}
  \begin{figure*}[ht]
  \begin{center}
  \includegraphics[scale=0.50]{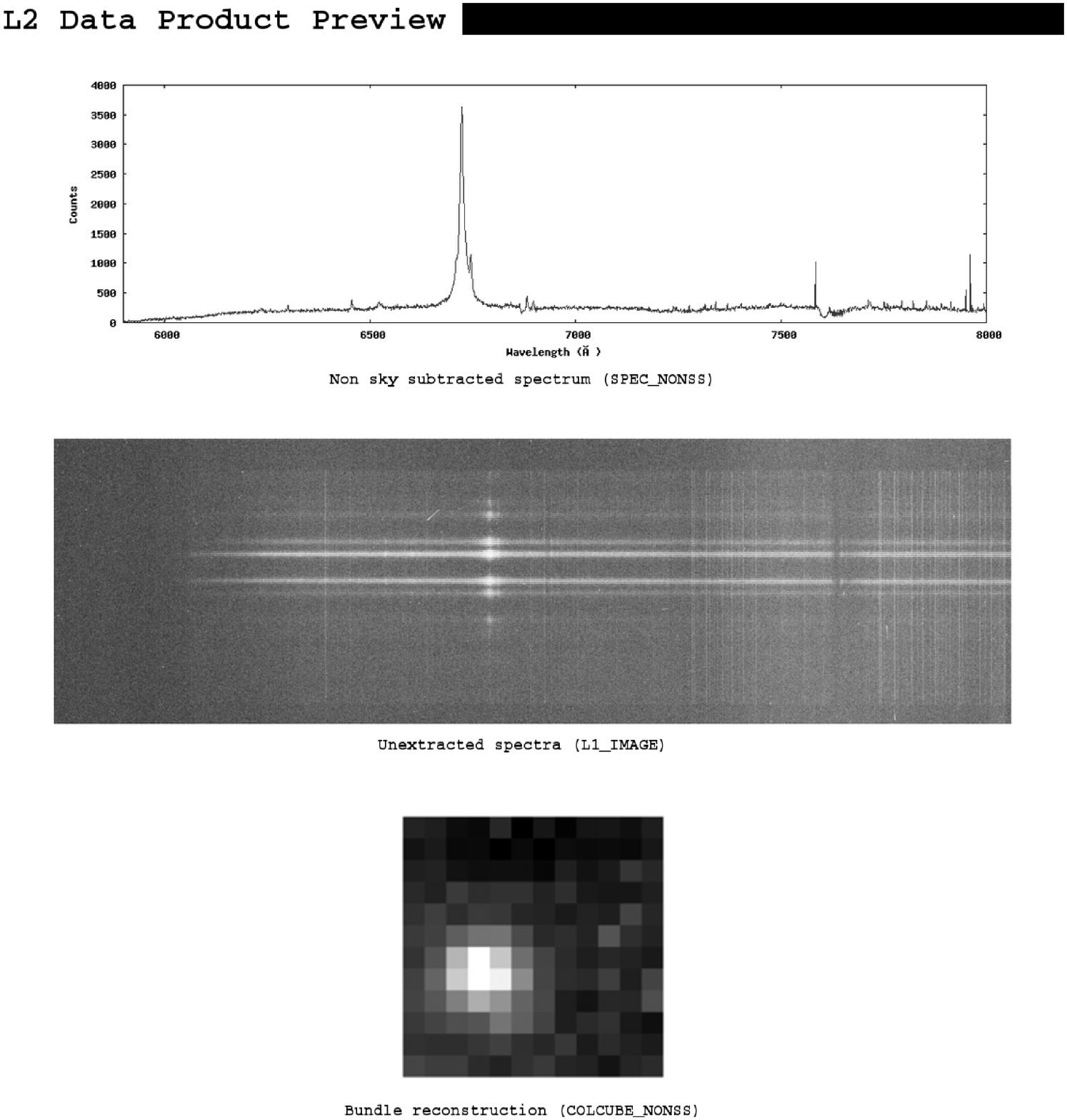}
  \caption{Example L2 reduction preview image.}
  \label{fig:preview}
  \end{center}
  \end{figure*}

  An example L2 reduction preview image is shown in Figure \textbf{\ref{fig:preview}}.

  \section{FRODOSpec L2 Reduction Pipeline Flow Chart}
  \begin{figure*}[ht]
  \begin{center}
  \includegraphics[scale=0.9]{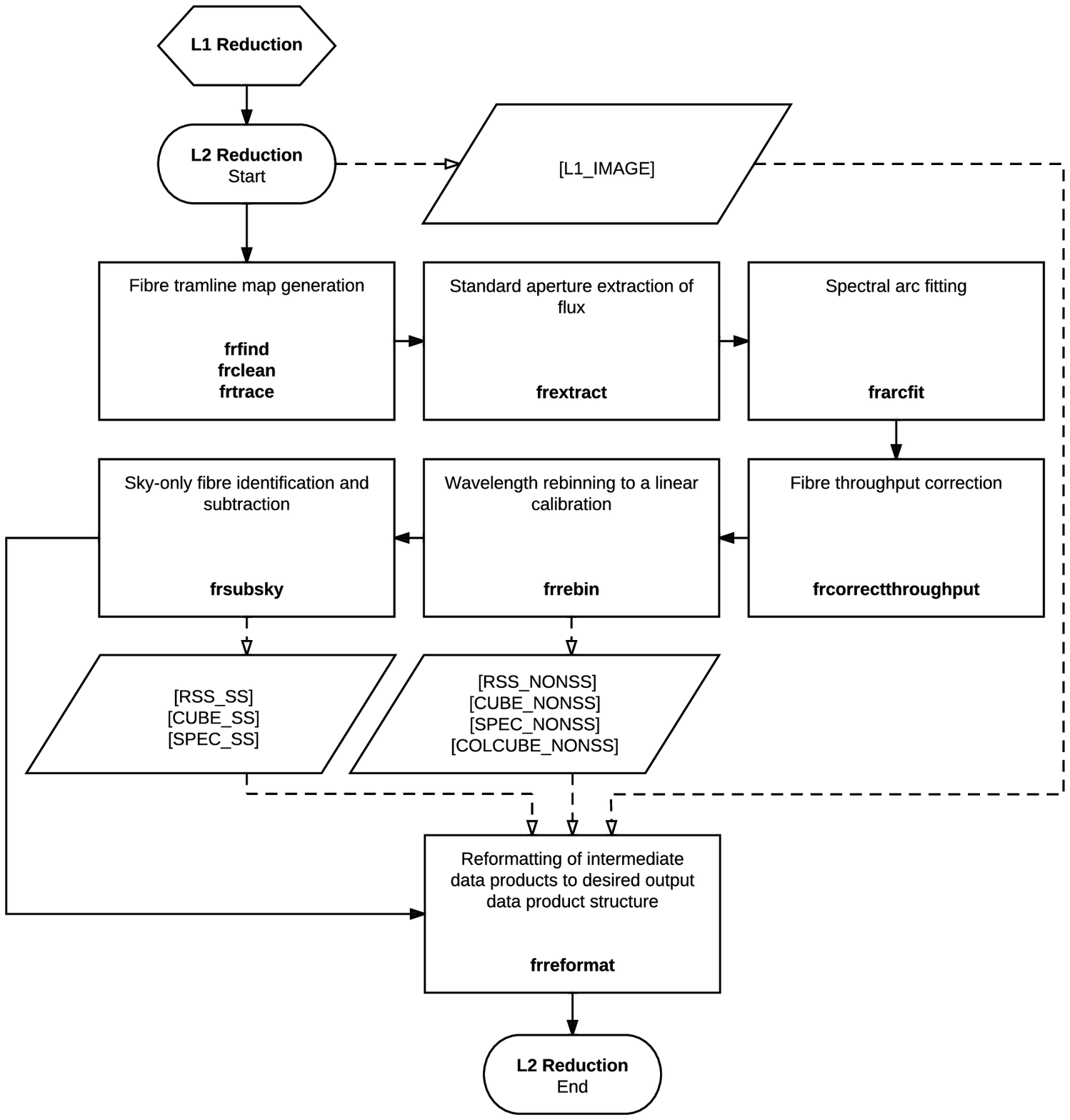}
  \caption{FRODOSpec L2 reduction flow chart. The processing sequence is shown with solid arrows. Processes that generate an intermediate data product that is 
  used to build the output science-ready data product are signified by a dashed arrow.}
  \label{fig:FrodoL2}
  \end{center}
  \end{figure*}

  A flow chart showing the processes and intermediate data products of the FRODOSpec L2 reduction is shown in \textbf{\ref{fig:FrodoL2}}.

}

\end{document}